\documentclass{collective-intelligence}

\usepackage{etoolbox}
\newtoggle{coins}
\toggletrue{coins}

\newtoggle{anom}
\togglefalse{anom}

\usepackage[utf8]{inputenc}
\usepackage[english]{babel}

\usepackage{calc}
\usepackage{natbib}
\usepackage{url}
\usepackage{listings}
\usepackage{hyphenat}

\usepackage{supertabular,array}
\usepackage{booktabs}
\usepackage{graphicx}
\usepackage{lipsum}
\usepackage{amsmath}
\usepackage{booktabs}
\iftoggle{coins}{}{
\usepackage{setspace}
\usepackage{txfonts}
}
\usepackage{placeins}
\usepackage{subcaption}

\usepackage[normalem]{ulem}

\usepackage[noblocks]{authblk}

\iftoggle{coins}{}{
\bibpunct{[}{]}{;}{n}{,}{,}
}

\begin{document}

\iftoggle{coins}{}{
\pagenumbering{roman}
\pagestyle{empty}

\thispagestyle{empty}
}

\selectlanguage{english}

\iftoggle{coins}{

\numberofauthors{6}

\iftoggle{anom}{
\author{
\alignauthor Mr. Scientist
\affaddr{University}\\
\affaddr{Place}\\
\email{scientist@university.edu}
\alignauthor Mr. Scientist
\affaddr{University}\\
\affaddr{Place}\\
\email{scientist@university.edu}
\alignauthor Mr. Scientist
\affaddr{University}\\
\affaddr{Place}\\
\email{scientist@university.edu}
\alignauthor Mr. Scientist
\affaddr{University}\\
\affaddr{Place}\\
\email{scientist@university.edu}
\alignauthor Mr. Scientist
\affaddr{University}\\
\affaddr{Place}\\
\email{scientist@university.edu}
\alignauthor Mr. Scientist
\affaddr{University}\\
\affaddr{Place}\\
\email{scientist@university.edu}
}
}{
\author{Petteri Räty}
\author{Benjamin Behm}
\author{Kim-Karol Dikert}
\author{Maria Paasivaara}
\author{Casper Lassenius}
\affil{School of Science, Aalto University, FI-0076 Aalto, Finland, first.last@aalto.fi}
\author{Daniela Damian}
\affil{University of Victoria, Victoria, BC, Canada, danielad@cs.uvic.ca}
}
\title{COMMUNICATION PRACTICES IN A DISTRIBUTED SCRUM PROJECT}
}{
\title{Communication Practices in a Distributed Scrum Project}

\author{Petteri Räty
	\and
	Benjamin Behm
	\and
	Kim-Karol Dikert
}
}

\maketitle

\iftoggle{coins}{}{
\thispagestyle{empty}

\newpage

\onehalfspacing
}

\begin{abstract}
While global software development (GSD) projects face cultural and time
differences, the biggest challenge is communication. We studied a distributed
student project with an industrial customer. The project lasted 3 months,
involved 25 participants, and was distributed between
\iftoggle{anom}{
the University X, North America and Y, Europe (anonymized for review).
}{
the University of Victoria, Canada and Aalto University, Finland.
}
We analyzed email
communication, version control system (VCS) data, and surveys on satisfaction.
Our aim was to find out whether reflecting on communication affected it, if
standups influenced when developers committed to the VCS repository, and if
leaders emerged in the three distributed Scrum teams. Initially students sent
on average 21 emails per day. With the reduction to 16 emails, satisfaction
with communication increased. By comparing Scrum standup times and VCS activity
we found that the live communication of standups activated people to work on
the project. Out of the three teams, one had an emergent communication
facilitator.

\end{abstract}

\iftoggle{coins}{}{
\newpage

\tableofcontents

\newpage

\pagestyle{plain}
\pagenumbering{arabic}
}

\iftoggle{coins}{
\newcommand{\OurScale}{0.25}
}{
\newcommand{\OurScale}{0.75}
}

\newcommand{\Finnish}{\iftoggle{anom}{European}{Finnish}}
\newcommand{\Finns}{\iftoggle{anom}{Europeans}{Finns}}
\newcommand{\Canada}{\iftoggle{anom}{North America}{Canada}}
\newcommand{\Canadian}{\iftoggle{anom}{North American}{Canadian}}
\newcommand{\Canadians}{\iftoggle{anom}{North Americans}{Canadians}}
\newcommand{\Victoria}{
	\iftoggle{anom}{ University Y }{ the University of
Victoria }}

\section{Introduction} \label{sec:introduction}

Global software development (GSD), meaning projects using teams in several locations, are
increasingly common. They allow geographical distances to be ignored and provide
access to more talented and skilled resources \citep{Herbsleb:2007}. Modern
software projects are often multi-disciplinary, which makes it hard to find
experts with required skills within one location \citep{Cavrak:2012}. In
addition, business globalization and market demand pushes companies to adopt
global software development practices, as they are forced to improve
time-to-market by adopting round-the-clock development \citep{Herbsleb:2001A}.

Agile methods are a collection of related iterative and incremental software
development processes. Agile methods were originally designed to improve communication by emphasizing
face-to-face communication \citep{Schwaber:2002}. This means agile methods have
traditionally been applied in a localised context. Scrum is one of these
methods which has its focus on day-to-day project management practises
\citep{Schwaber:2002}.

A global aspect in a project decreases the amount of face-to-face communication
making communication more complicated \citep{Monasor:2010}. The overall amount
of communication and collaboration will decline as good communication gets
harder to achieve because of a geographical separation \citep{Herbsleb:2007}.
The lack of personal contact
impedes the ability to build understanding and trust among parties
\citep{Jarvenpaa:1998}.
There is ongoing research of ways to mitigate these problems in GSD, and agile
methods adapted to GSD are rising as a candidate solution \citep{Paasivaara:2013}.
It is necessary to substitute live communication
with technical aids such as videoconferencing, email, and chat, which helps to
maintain especially informal communication.

The interest in GSD teaching and education will increase, as it is becoming a
common practice in the software industry and therefore there is a growing demand
for knowledge \citep{Monasor:2010}. \citet{Cavrak:2012} mention that GSD projects
face not only technical problems, but also difficulties in handling both
cultural and language differences. They also show that the lack of trust and
cooperation can be a consequence of cultural misunderstanding, which is usually
caused by weak communication \citep{Cavrak:2012}. For this reason, it is highly
valuable to teach working in a distributed environment and to encourage students
to cooperate with people from different cultures, and at the same time prepare
students to work in a global environment after graduation.

In this research, we investigate communication practices and perceived
satisfaction in an industrial student project arranged in cooperation with two
universities. We study how students changed used communication practices based
on their reflection in a project where all team members were not able to meet
each other face-to-face. In addition, we inspected what kind of effect standups
(short Scrum mandated status meetings)
had on the commit activity of the students, and whether leaders emerged in Scrum
teams. Based on the results, we have three proposals for future GSD projects.
The first is that developers are made aware of the problems involved in sending
too many emails. The second is to emphasize the importance of standup meetings,
not only as a means of sharing information, but also as a way of increasing
motivation to work on a project when there are other projects going on at the
same time. The third is to use communication visualization and metrics as a tool
in retrospectives so that teams are more aware of their communication practices
and can use the information as a base for their potential process changes.

The paper is structured as follows. We start with the background of the research
including explanations of GSD and Scrum. Then the section Research Design
 shortly describes the course, what data we inspected,
and clarifies what kind of data we had and how it was analyzed. After going through
results, in Discussion we summarize our findings and discuss a little
about possible limitations of the research. Finally
we conclude our findings and propose a direction
for future work.

\section{Background} \label{sec:background}

In this section we describe GSD in general and how agile methods can be applied
in GSD. We pay particular attention to the agile method called Scrum and how it can be applied in GSD, as Scrum was used in the
case project. As a comparison with our case
study we present a brief review on how GSD skills have been taught in other
courses.

\subsection{Agile methods in global software development}

Agile software development has its foundation in enabling change
\citep{Dingsoyr:2010}. The ability to cope with change hinges on the internal and external
collaboration and communication of development teams. Delivering working
software in short iterations, and integrating and testing the product
continuously are important practices in agile development \citep{Highsmith:2001}.
These practices aim to provide continuous feedback and transparency, which
improves control and the ability to cope with change.

Global software development is a well established trend, which aims to gain
benefits by distributing development effort across multiple sites. The main
reasons to make development global are expected cost savings as well as the need
for extra people and knowledge \citep{Smite:2011}. The benefits are however
weighted against challenges, which in GSD are due to the difficulty of arranging
control over a distance \citep{Herbsleb:2007}. A centralized development setting
allows frequent interactions and eases creating a shared understanding, but a
distributed setting lacks these benefits. The main challenges of GSD are less
effective communication, which results in a lack of understanding and awareness
of what people at other sites are doing, and incompatibilities in ways of
working and the use of tools \citep{Herbsleb:2007}.

Agile and GSD seem to have a very different basis, as agile emphasizes close
collaboration, whereas in GSD collaboration is hampered by the necessity to work
across organizational and geographical boundaries. Nevertheless, agile practices
are used in GSD, and a potential is seen for combining the benefits of both
disciplines \citep{Hansen:2011}. As these two disciplines have different
fundamentals, it is clear that problems arise, with communication being the
foremost \citep{Paasivaara:2006}. Agile methods emphasize frequent face-to-face
communication, which may be difficult to emulate between remote sites. However,
striving to satisfy this requirement may also be seen as a strength as enabling
frequent communication is a recommended practice in GSD \citep{Smite:2011}. Other
challenges are the requirement for short iterations and frequent integration,
but when implemented properly agile methods provide transparency and control
\citep{Paasivaara:2006}. Agile methods are often seen as sets of simple rules,
which should be used as a basis for an organization to learn and build their own
optimal way of working \citep{Highsmith:2001}. This implies that it is meaningful
to apply agile methods partially, even if the environment is not on their home
ground, such as in a distributed project. Selectively applying suitable agile
practices has proven to benefit distributed projects \citep{Holmstrom:2006}.

\subsection{Scrum in GSD}

Scrum is an agile software development method, which has its focus on day-to-day
project management practices. The main elements of Scrum are splitting work into
prioritized increments, delivering working increments of software in time boxed
sprints, daily standup meetings to keep everyone on track, and enabling teams to
self-organize \citep{Schwaber:2002}. Work is divided in to 2-4 week sprints. The
sprints start with a planning meeting to create the initial scope. At the end of
the sprint there is a demo where the work results are demonstrated to
stakeholders. The ending of the sprint also includes a retrospective session in
which the team can adjust the process to solve problems they have observed.
Scrum emphasizes empirical control in order to react to change
\citep{Schwaber:2002}. Empirical control implies frequent communication which is
manifested in Scrum as standup meetings and a preference for face-to-face
communication.

% [Hossain2009 --> Scrum and GSD] --> What problems emerge? --> What problems %
% are solved? --> How does it fit?

\citet{Hossain:2009} found in their systematic review of literature that communication issues
are major challenges when using Scrum in distributed development.
The possible communication
challenges and misunderstandings could be partially avoided by providing
synchronous communication. This can be arranged, for example, using synchronized
working hours or generating Scrum teams that are not dependent on offshore
teams. Team members' cultural background has been shown to have an impact on collaboration and
communication. Furthermore, coordination may also arise as a challenge if Scrum
teams do not have proper facilities to support the needs of daily communication.
In addition to these, offshore team members might feel discouraged to voice
their opinion aloud, which might cause confusion and misunderstanding among
people on other site, but is actually caused by cultural and linguistic
disparity. \citep{Hossain:2009}

\subsection{GSD courses}

% --> Viewpoints on Scrum courses, are they common? % --> See [Paasivaara2013]
%(ref. 7, 8) on talk about GSD on courses?

During the last decade, the amount of publications related to teaching GSD has
increased considerably, which indicates emergent interest in the subject
\citep{Monasor:2010}. \citet{Swigger:2012} looked at when distributed student
teams work and found that students often work outside the normal workday.
\Citet{Hossain:2009} found that there is a growing trend of papers covering the
subject of using Scrum practices in GSD context. Still most reported GSD courses
have used the waterfall process instead of agile methods, which reflects the
contradiction between education and industry \citep{Paasivaara:2013}. As a clear
link to the success of Scrum practices in GSD is missing, it is important to put
more effort on studying this area in practice. Agile methods have a clear place
in GSD and their impact could be emphasized more on GSD courses.

% % --> As agile is reported to be taught quite little, it is good to have this
% study. % --> Why the course was arranged in general? --> Some reason why this
% course is as it is?

\section{Research Design} \label{sec:method}

In this section, we first describe our case study project. We then present our
research questions and the related hypothesis. Finally we list the collected
data and briefly explain how it was analyzed.

\subsection{Course description} \label{sec:case_study}

The research was based on a case study following \citet{Yin:1994ad}. The case
study was arranged as a collaborative course between
\iftoggle{anom}{
the University X, North America and Y, Europe (anonymized for review)
}{
the University of Victoria, Canada and Aalto University, Finland
}
in spring 2012. The course aimed to teach
students important GSD skills by constructing a real software development
environment using agile methodologies \citep{Paasivaara:2013}. To accomplish this,
the course coordinators had procured a real customer, for whom the software was
developed.

There were some minor differences in the course settings at both locations. The
\Finns{} started their course in September 2011, as a part of a Software
Development Project course. During the fall, students got familiar with the
Scrum framework and started using it as a development process. They were
instructed to use distributed Scrum when the \Canadian{} team joined them in
January 2012. The students worked together until the course ended at the
beginning of April, even though the \Finnish{} course officially ended at the end
of February. After the \Canadians{} joined on the course, all students excluding a
Scrum Master were divided in to three teams, so that each team consisted of 7 to
8 students and contained students from both countries. The core idea was to
implement a more accurate Scrum environment in comparison to earlier research groups
such as \citet{Scharff:2012}. The Scrum Master (also the first author) was
shared between the three teams because he was the only one with previous
experience with the source code and had industry experience working in that role. This enabled all teams to tap into his knowledge as needed.

The \Canadian{} instructor had selected a student per team for a role that was
described before the project started as a team leader.
Because there are no leaders in Scrum, it was
emphasized to all students that the role is to serve as the contact person
between the \Canadian{} instructor and the team so that the \Canadian{} instructor has
to only deal with one point of contact. However, we can not rule out the
potential effect this had on the students.

The cultural background of the students in the \Finnish{} location was uniform as
all the students were \Finnish{}. However, in \Canada{} many students had grown up
elsewhere. In addition to \Canadians{}, the course at \Victoria
had students from other countries, for example Iran and China. Furthermore,
English was the language of communication, which was not a native language for
twelve (vs. 11 natives) of the students. In both locations the students were expected to spend
about 10 to 12 hours per week on the project. For \Finns{} this was their full
amount of hours for the project but for the \Canadians{} the course also included
lectures, reading and blogging. The experience background in programming for the
students varied from taking a single basic programming course to having a decade
worth of experience. When creating the teams, the students and instructors aimed
to create three teams that were as equal in skill and experience as possible.

Students used both synchronous and asynchronous communication. Google+ Hangout
was primarily used for videoconferencing in Scrum meetings. Skype was also used
to for example reach the product owner. In addition, email was used for a
project or team related asynchronous communication, and internet relay chat (IRC) for instant messaging
inside teams. IRC was a familiar tool to \Finns{} but \Canadians{} had less previous
experience with it. The course provided each team and the whole project mailing
lists that everyone was allowed and encouraged to use in order to reach a
suitable audience.

For our research and educational purposes there were mandatory practices. The
sprints were of equal length and lasted two weeks. Having the sprints of equal
length makes it easier for us to compare the sprints for research purposes, and
for education purposes enables learning the Scrum practices by repetition. The
students were required to have two Scrum standup meetings per week. Normally the
meeting happens daily in a Scrum team working full time in an industrial
setting. Because the students only worked part time, having two standups per
week allowed roughly the same amount of work to happen between the standups. The
students were allowed to agree among themselves for the time of the standups
because neither location had time dedicated in the course schedule that could be
used for that purpose. At the beginning of the project we selected Google+
Hangout as the videoconferencing tool because it allowed the needed amount of
simultaneous connections and was available for free on all operating systems.
During the project Google also released mobile applications allowing students to
connect to the standups from the road. The version control system used in the
project was Git.

\subsection{Research questions}

The goal of our research was to find out whether reflecting on communication had
an effect on the actual communication, if standups influenced the time when
developers made commits to the VCS repository, and if leaders would emerge in
the teams.

\emph{Does reflecting on communication affect the actual practices of
communication?}

In Scrum, teams should be self-organized and thus decide the best practices by
themselves \citep{Schwaber:2002}. Based on this we hypothesized that teams would
try to adjust used communication practices based on the reflection of previously
used practices. Teams were asked to keep a retrospective after each sprint in
which team members went through the current sprint and discussed problems they
had observed. The goal was to identify improvements and implement them in the
next sprint. Hence, it would be logical to see communication practices change
whenever people were not satisfied with the way they are used.

\emph{Do standups activate developers to do work?}

The students did not have a set schedule to work on the project and from
previous work we know that students work irregular hours \citep{Swigger:2012}.
Our hypothesis is that the live communication of standups activates students to
commit more if there is a standup near. In the standup you need to have answer
the question: ``What have you done since the last standup?" The social pressure and the fact that you must spend at least the standup thinking about the project should mean that students commit more near standups.

\emph{Do leaders emerge in the distributed Scrum teams?}

There are no designated leaders in Scrum teams. We wanted to study the
communication of the teams to see if developers emerged that could be
characterized as dominant communicators due to their communication patterns. We
know that leaders can have a different ratio on how many messages they send and
receive \citep{Gloor:2003} and we can use that knowledge to see if such patterns
emerge in this context.

\subsection{Collected data}

Data was collected using techniques listed in table \ref{table:
datacollection} \citep{Paasivaara:2013}. The most important
data sources were sprint surveys, sent and received email messages, and version
control data. At the end of every sprint, a web-based sprint survey was sent to all 25 students, and 19-22 submitted surveys for a response rate between 76\% and 88\%. The surveys contained 23 stationary questions with a total of 101 sub-questions emphasizing the perceived satisfaction, trust, used tools, and usefulness of Scrum and communication practices. The surveys were conducted for our scientific purposes
and are not a mandated part of the Scrum process but they do also force the
developers to reflect on the experiences from the current sprint. The emails
were collected by having the students submit the email headers from all the
course related messages to the course personnel at the end of the project. This
way the students could be confident that the bodies of messages were never read
by the teaching personnel. All students were assigned random numeric user identifiers so in this paper u10 means developer 10. The three teams were assigned similar random numeric identifier 0, 1, and 2.

\begin{table*}[!hbt] \caption{Data collection}
\centering
\footnotesize
\centering
	\begin{tabular}{ p{4cm} p{5cm} p{5cm} l }
		\toprule
		\textbf{Collection} & \textbf{Purpose/Instruments} &
		\textbf{Data collection} & \textbf{Analysis} \\ \midrule
		Pre-course survey & Student background, expectations & 25 responses	&
		 \\
		Sprint surveys & Trust, Teamwork, Communication &
		19-22 responses& \\
		Email & Communication & 4210 emails & SNA \\
		Chat & Communication & 2859 messages & SNA \\
		Meeting videos & Communication & 31 Daily Scrums, 2 Demos \& Sprint
		Planning meetings & \\
		Observations & Teamwork, communication & 5 Demos \& Sprint Planning
		meetings, Retros, Daily Scrums, teamwork in war rooms & \\
		Student logs & Communication, encountered issues, tasks & & \\
		Agilefant & Estimation accuracy, work breakdown & Task estimates and
		actuals, burndown & \\
		Post-course survey & Learning & 20 responses & \\
		Interviews & Learning, communication, communication tools, improvement
		ideas \& 13 interviews, 40-90 min each (8 \Canadian{} \& 4 \Finnish{}
		students, \Finnish{} PO) & & Atlas.ti \\ \bottomrule
	\end{tabular}

\label{table: datacollection} \end{table*}

\iftoggle{coins}{}{
\FloatBarrier
}

\subsection{Data analysis}

Data analysis was primarily conducted using quantitative analysis supported with
qualitative observation. To analyze the data centrally all the data was
collected to one relational database. The survey data was primarily analyzed
using R programming language, which produced diagrams that we used for
qualitative analysis of how students' opinion changed during sprints. The same
method was used for VCS data analysis. For communication, we studied the amount
of sent and received emails per a person in order to analyze students' activity
and how communication practices changed during sprints.

\section{Results}
\label{sec:results}

In this section, we present the results of our data analysis. This section is
divided into subsections for each research question. A central part are the
figures that resulted from our analysis.

\subsection{Communication and Satisfaction}

\begin{figure}[h]
\centering
\includegraphics[width=0.8\linewidth]{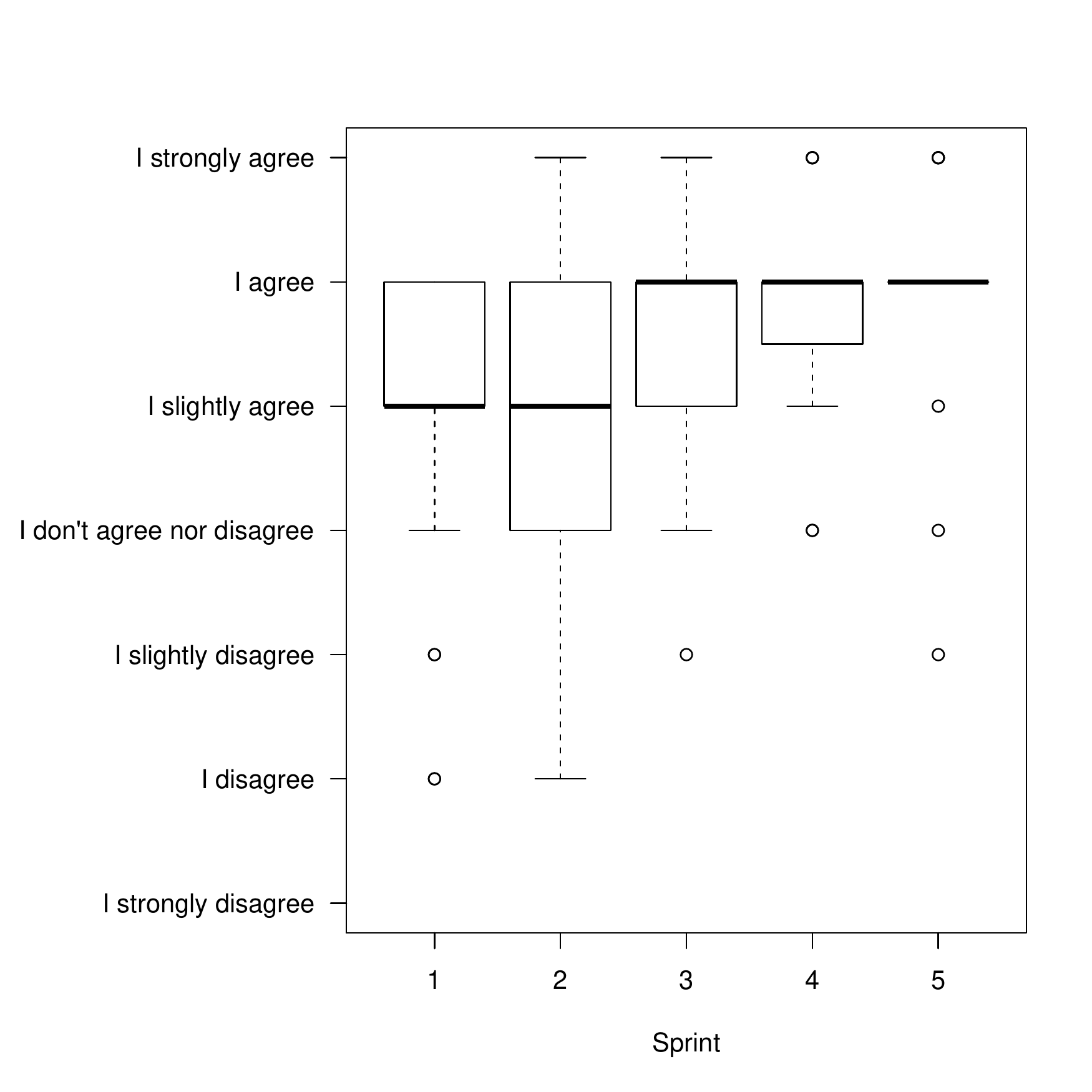}
\caption{Communication satisfaction survey answers}
\label{fig:communication_satisfaction}
\end{figure}

The answers to the question: "I am satisfied with the communication practices
that are used in my team." are shown in
Figure~\ref{fig:communication_satisfaction}. The boxplot, as such, shows the
students' satisfaction for the used communication practices. The plots for the
all the individual teams followed the same trend where they were least satisfied
in either sprint 1 or 2 and the satisfaction increased towards the end of the
project. We can also observe that while the median stays the same in sprints 3
to 5, the interquartile range gets smaller. This means that more and more
respondents became satisfied with the communication practices as the project
progressed.

\begin{figure}
\centering
\includegraphics[scale=\OurScale]{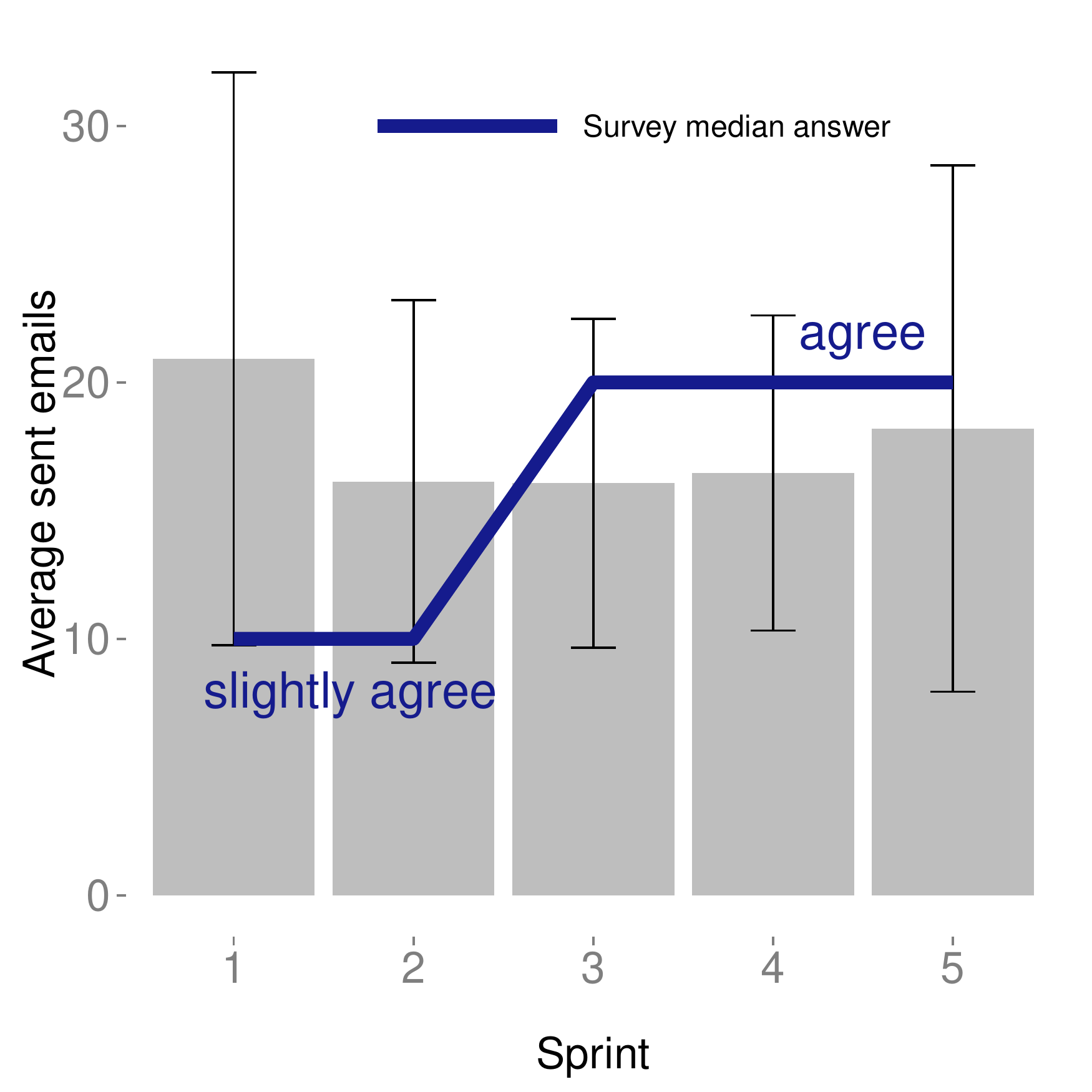}
\caption{Email variation and satisfaction}
\label{fig:email_variation}
\end{figure}

In Figure~\ref{fig:email_variation}, we have plotted the average sent emails by
developers with the corresponding standard deviations superimposed with the
median information from Figure~\ref{fig:communication_satisfaction}. In this
text we report the number rounded to the nearest integer. The average amount of
daily emails was 21 in sprint 1 with a standard deviation of 11 emails. In
sprints 2-4 the email count was 16. The variation decreased to 7 in sprint 2 and
to 6 in sprints 3 and 4. For the final sprint the number of emails increased
again to the level of 18 emails with a variation of 10. The changes to the final
sprint can be explained by the activities involved in ending the project. In our
observation we noted that multiple students said that they were getting so many
emails that they started skipping reading them.

% \begin{enumerate}
% 	\item A good quote from Maria about email flood
% 	\item Does the change in satisfaction just happen naturally?
% \end{enumerate}

\subsection{Commits and Standups}

As mentioned in Course Description, each team had two standups per
week that they were allowed to select the times for on their own. For \Victoria
23 standups happened during the evening and 27 standups happened during the
morning. This means that almost as many standups happened during the evening as
the morning.

We restricted our analysis to the 30 hour window before and after standups. Had
we selected a 40 hour time window, it would have skewed the results. This is due
to the fact that there were only two standups per week, and we know that students worked all days of
the week including the weekend. Normally the closest standup is about a day
behind or ahead but this is not the case over the weekend. This means after 30
hours there are less working hours during the sprint so the number of commits
cannot be directly compared, hence the time window.

\begin{figure}[h]
\centering
\includegraphics[scale=\OurScale]{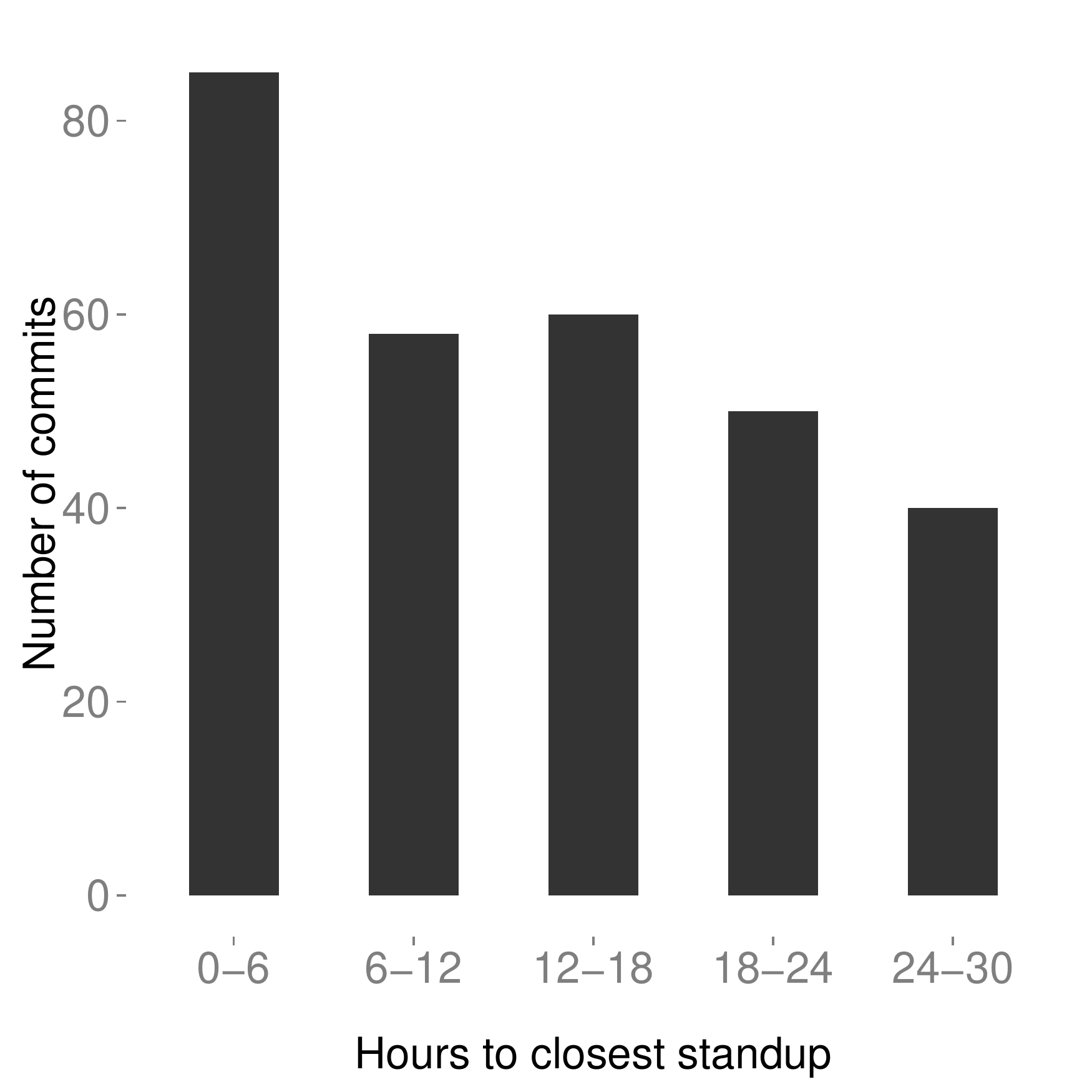}
\caption{Commits in relation to standups}
\label{fig:commits_standups}
\end{figure}

In Figure~\ref{fig:commits_standups}, we have plotted the relation of Git
commits to the nearest standup. We grouped the commits together to six hour
windows. We selected six hours so that each window got tens of commits so that
specific events had less influence. This means that a commit done 2 hours before
as well as a commit done 3 hours after a standup would show in the bar with x-axis value 0-6. The relationship appeared linear therefore we ran Pearson
correlation. The correlation was high at -0.926 with a p value of 0.02. The
clearest finding here that is not influenced by time of day, is found by
looking at the first and last bars. Here the hours are exactly one day apart
and we can see that near the standup more than twice the amount of commits
happened. In the category of 0-6 hours to the nearest standup there are 85
commits and for 24-30 hours there are 40 commits.

\begin{figure}
\centering
\includegraphics[scale=\OurScale]{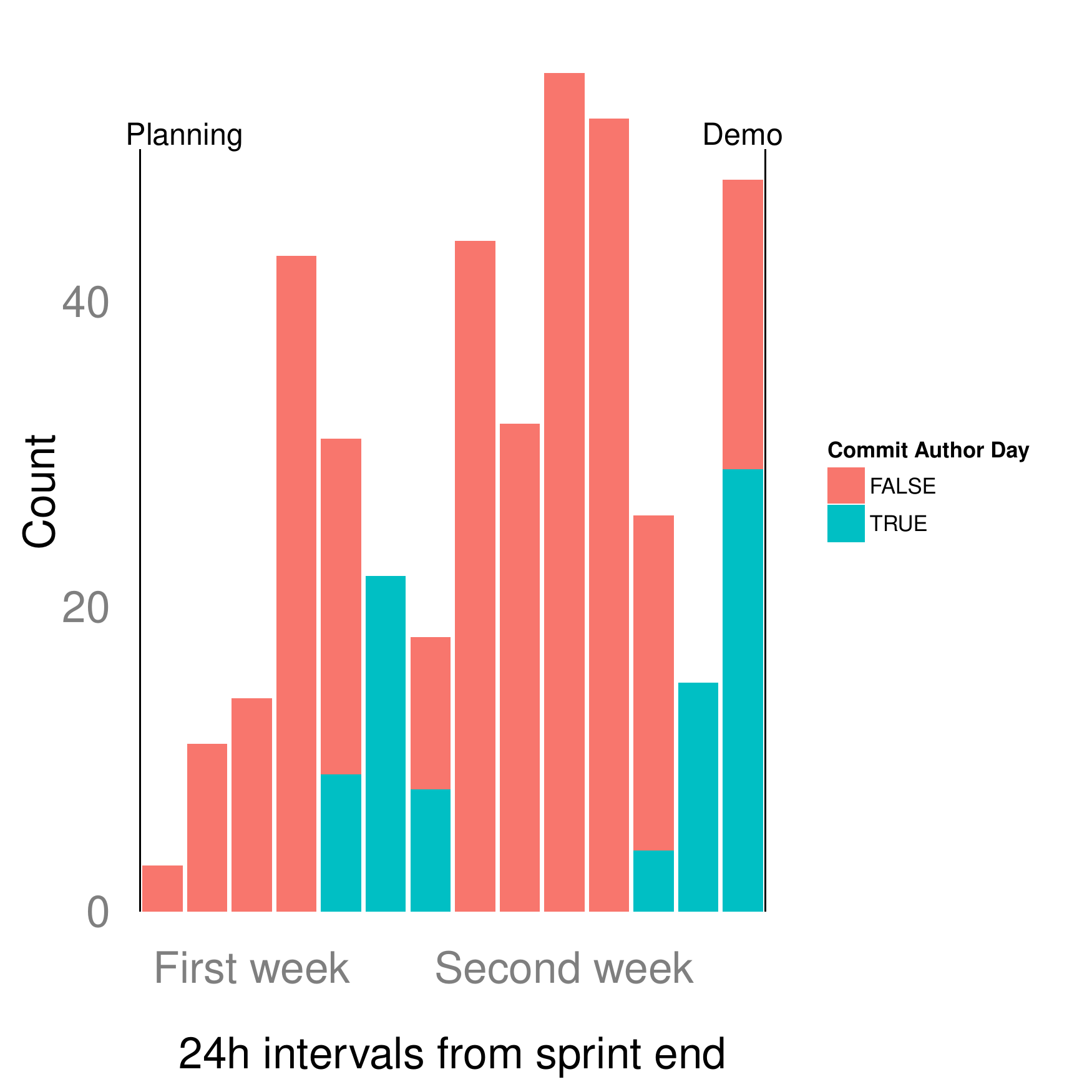}
\caption{Commits during sprints}
\label{fig:commits_during_sprint}
\end{figure}

\begin{figure}
\centering
\includegraphics[scale=\OurScale]{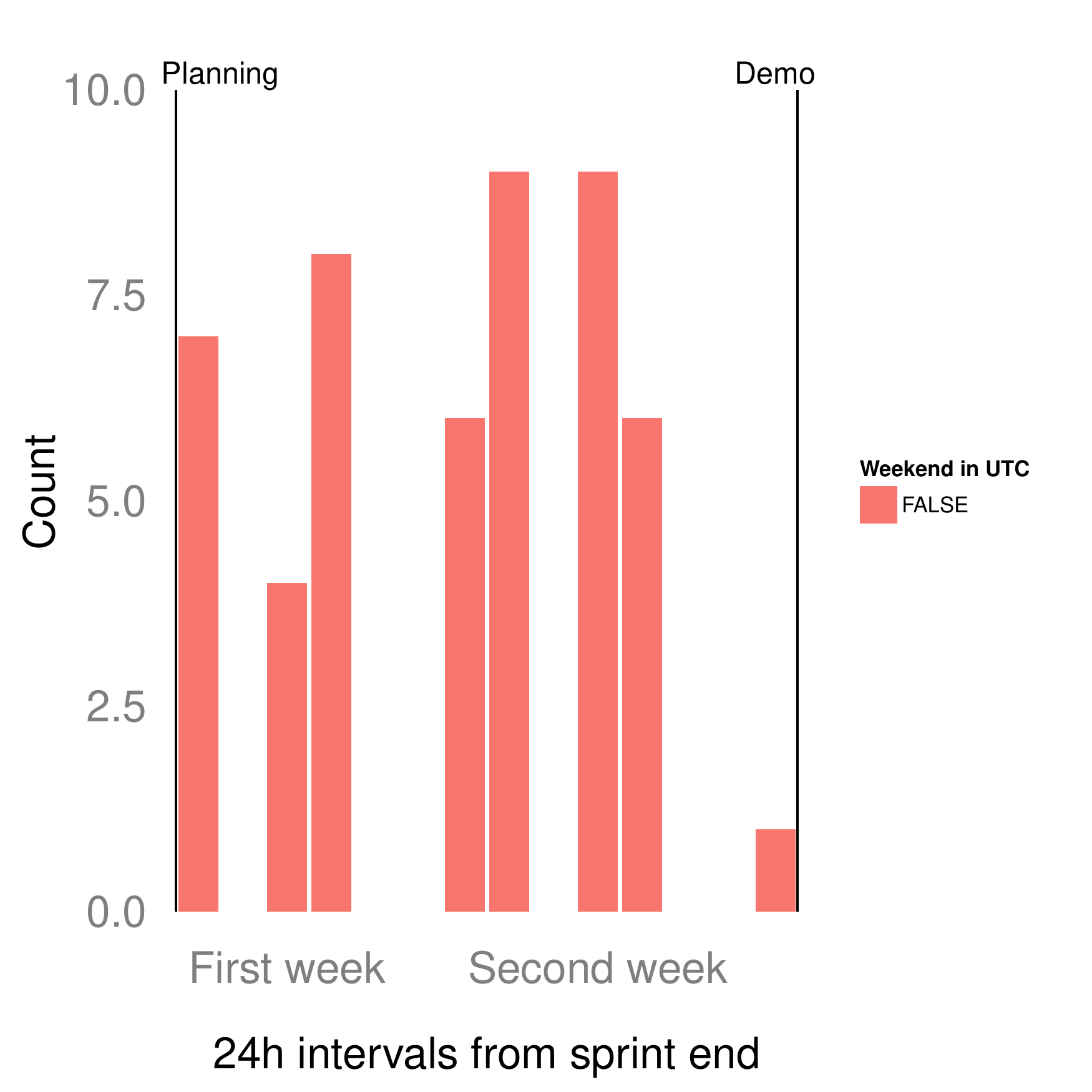}
\caption{Standups during sprints}
\label{fig:standups_during_sprint}
\end{figure}

In Figure~\ref{fig:commits_during_sprint}, we are
looking at the linear timeline of a sprint. On the x-axis, we have the number of 24h intervals left until the end of the sprint. As
mentioned, sprints end with the demo session, which happened on Mondays every
two weeks. The exception to this was sprint 2, which ended with the start of the
\Canadian{} reading break. From this figure we can see that students worked during the weekends in
addition to working during the week. We can see that people seldom committed
during the first day of the sprint. We can also observe that a lot of activity
happened at the very end of the sprint. Closer look into the distribution of these
commits reveals that it is mainly due to team 1 with a share of 69\%. Team 0
accounted for 25\% of the commits. In Figure~\ref{fig:standups_during_sprint}, we
have the same x-scale with the number of standups on the y-axis. We can see that
the first rise to over 40 commits coincides with the second standup of
the first week located at the fourth bar. Our interviews supported this fact. A
\Canadian{} student said:
\begin{quote}
If we could somehow have actually Daily Scrums, that would force people to at least look at their project every day.
\end{quote}

\subsection{Emergent Leaders}

\begin{figure}[h]
\centering

\begin{subfigure}[b]{0.3\textwidth}
\centering
\includegraphics[scale=\OurScale]{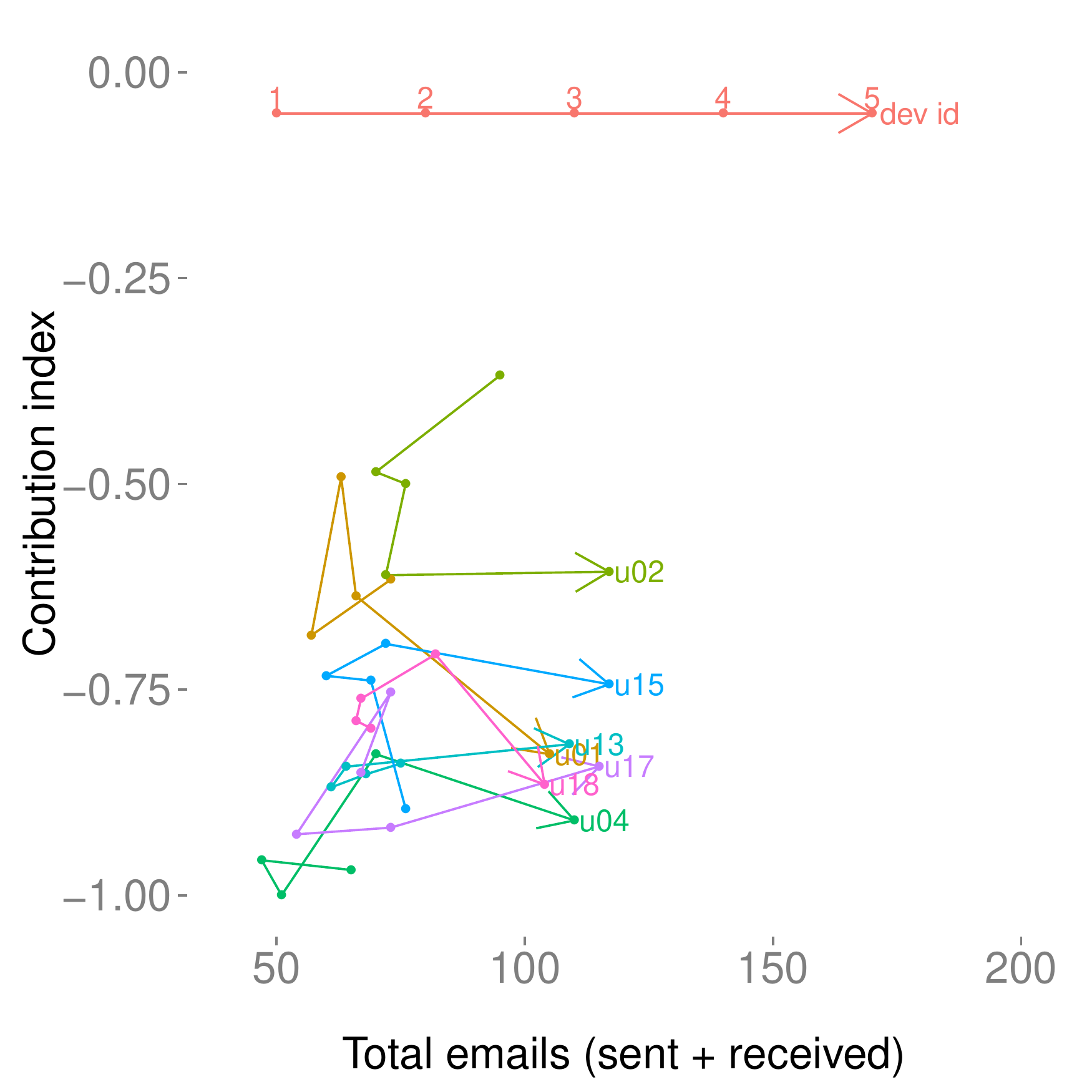}
\caption{Team 0}
\label{fig:cit0}
\end{subfigure}

\begin{subfigure}[b]{0.3\textwidth}
\centering
\includegraphics[scale=\OurScale]{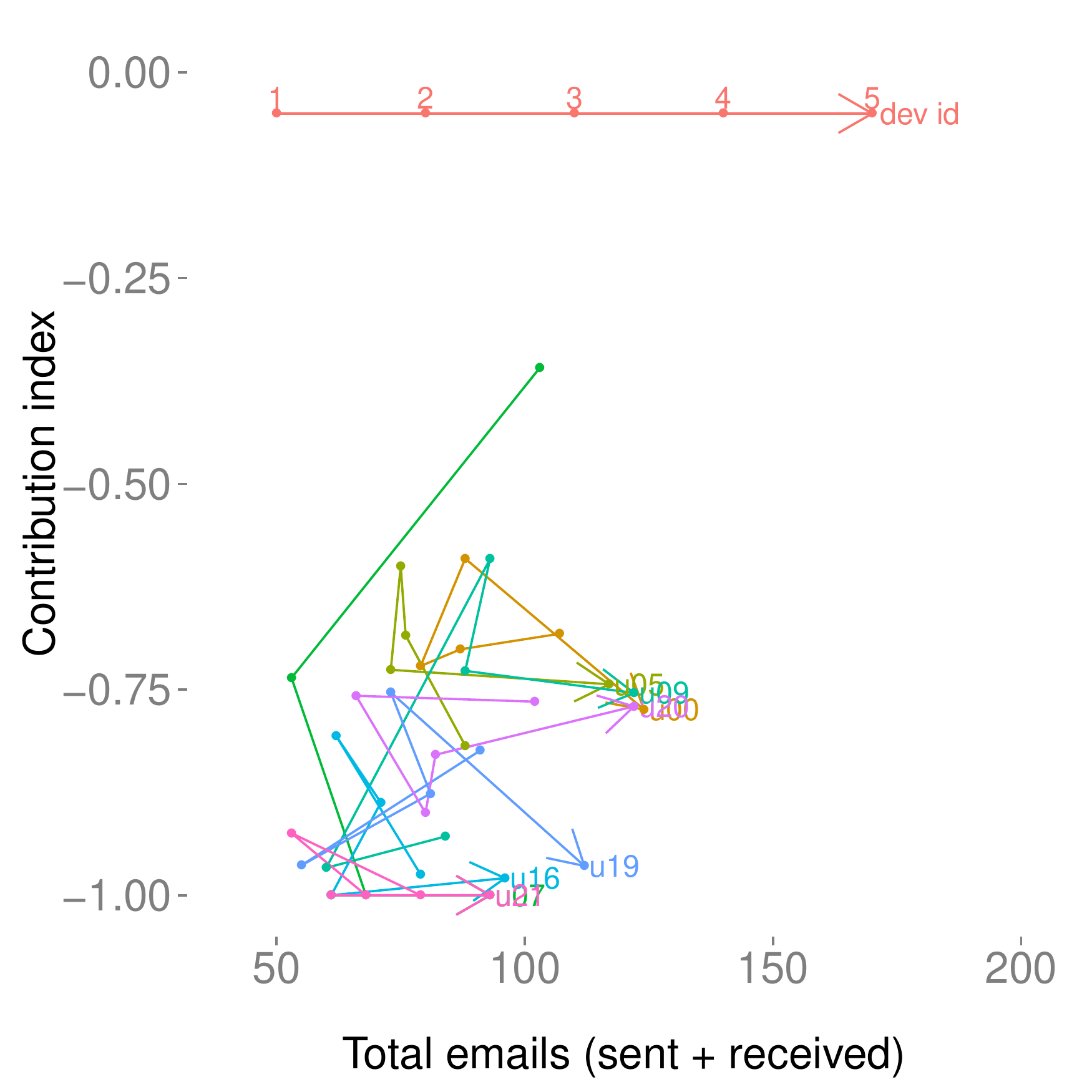}
\caption{Team 1}
\label{fig:cit1}
\end{subfigure}

\begin{subfigure}[b]{0.3\textwidth}
\centering
\includegraphics[scale=\OurScale]{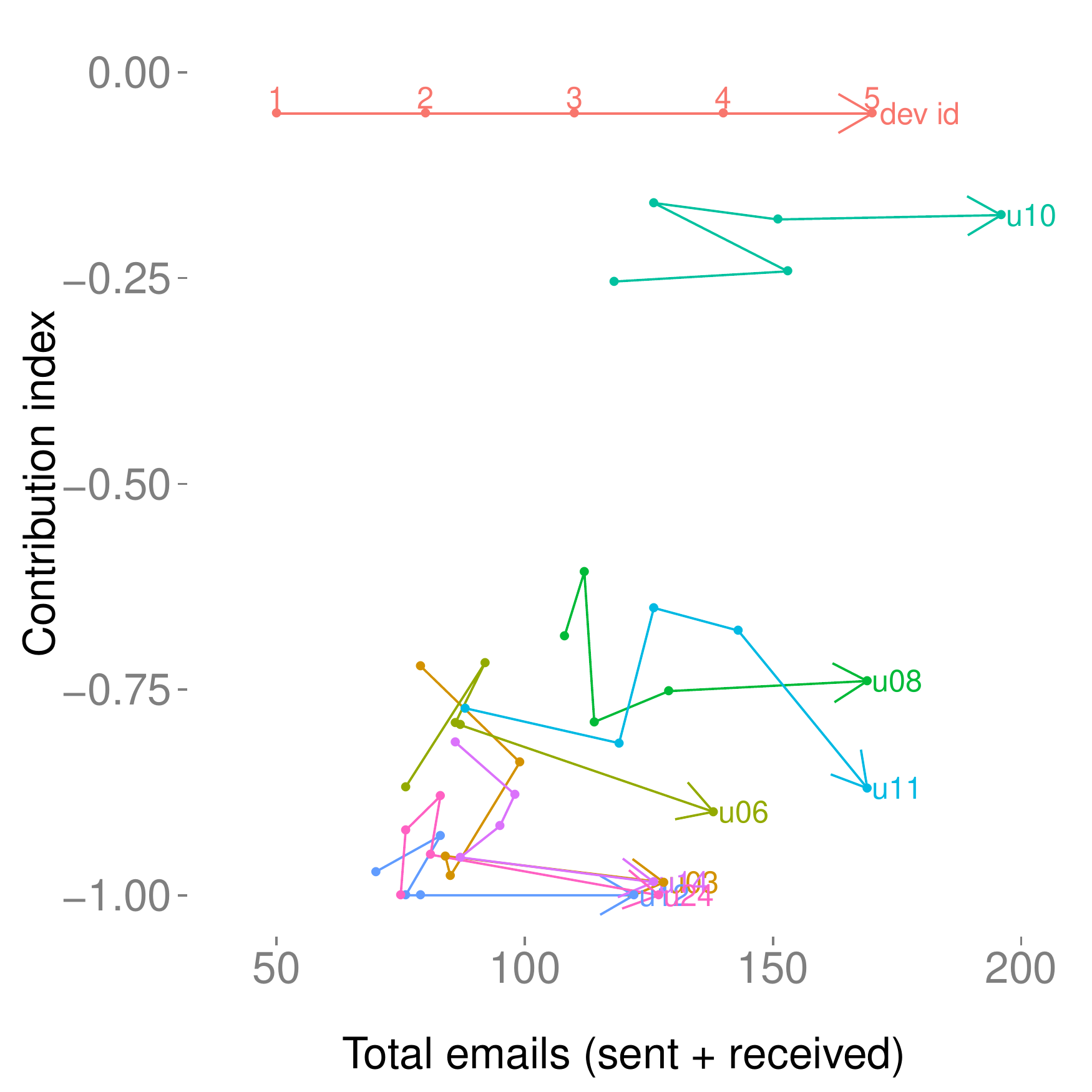}
\caption{Team 2}
\label{fig:cit2}
\end{subfigure}

\caption{Contribution indexes for sprints 1-5}
\end{figure}

We have the contribution indexes for all the different teams in Figures
\ref{fig:cit0}-\ref{fig:cit2}. Contribution index is defined as
\[\frac{\text{messages sent} - \text{messages received}}{\text{total of messages
sent and received}}\] \citep{Gloor:2003}. Leaders have been found to have their
contribution index near zero meaning that they sent and received an equal amount
of messages. For each developer the contribution index has been plotted per
sprint and the values connected with a line that has an arrow in the end. This
means that u02 in Figure~\ref{fig:cit0} had a total of about 120 messages in
sprint 5. Analyzing teams 0 and 1 in Figures \ref{fig:cit0} and \ref{fig:cit1}
respectively, we can see that while there are activity differences between
developers no single individual stands out. However in Figure \ref{fig:cit2} a
leader, u10, has emerged. U10 was also the team 2 coordinator with the
\Canadian{} instructor.

\iftoggle{coins}{}{
\FloatBarrier
}

\section{Discussion} \label{sec:discussion}

Now that we have presented our results, it is time to answer our research
questions based on the data. After that we discuss possible limitations of our
research.

\bigskip
\bigskip

\subsection{Communication and Satisfaction}

First, we wanted to answer the question: "Does reflecting on communication
affect the actual practices of communication?". By analyzing the actual email
traffic we saw that as sprints progressed the amount of emails sent daily
reduced by 20\% while satisfaction went up so that everyone in the interquartile
range agreed with the statement by the fifth and final sprint. At the beginning we observed that
the biggest communication problem was considered to be the flood of emails.
Developers thought that with the limited amount of time, too many emails
interrupted their focus on the main work, and saw this as unpleasant. For this
reason, students preferred synchronous communication over emails
\citep{Paasivaara:2013}. As daily email flow was reduced, developers were more
satisfied. We speculate that had the project continued with more sprints the
profile for sprint 5 would have been similar to sprints 2-4.

% In addition, synchronous communication was seen as better since it helped
% students to understand each other better and to increase team spirit.

By looking at the survey data and the actual amount of email communication we
were able to find a pattern of behavior. This pattern was quite similar to the
pattern that \citet{Swigger:2012} found in their study. They argued that
students tend to be more active communicators at the beginning of the project,
after the middle of the project, and at the end of the project. Their findings
match well with our findings of the pattern of sent emails, excluding the fact
that we did not notice a peak in the middle of the course in sprint 3. In our
case, the average amount of sent emails in sprint 3 was almost equal with
sprints 2 and 4.

\subsection{Standups and Commits}

Secondly, we looked at the relation of when Git commits are done to Scrum
standups. We found a high correlation between the two meaning that standups
heavily influence when students work. From previous work \citep{Swigger:2012} we
knew that students worked irregular hours and dates. In the process of our
analysis we were able to confirm those results and provide a further
contribution. The standups pace when students spent their budgeted time on the
project.

This information is useful to both teaching staff and industry settings where
people work part time. The original purpose of the standups is for the exchange
of information and to help solve problems within teams. In contrast to the
waterfall model that most courses use, having the standups pacing the work
means there is less danger of students trying to do most work at the end of the
project. We expect similar behavior in the industry for people who
have for example split their time between multiple customers.

%\begin{enumerate} \item A Juicy Quote About Standups \end{enumerate}

\subsection{Emergent Leaders}

Finally, by looking at contribution indexes and using knowledge about how it behaves from
previous works, e.g., \citep{Gloor:2003} we were able to see if leaders emerge in
our Scrum teams. One of the three teams clearly had a developer with a facilitator
role. We speculate that it would be useful for the teams to see these
analyses during the project to be able to reflect on the communication and
conclude if the patterns are good or bad for them.

\subsection{Other Practices}

In this research we have focused on the main textual and verbal communication
medias of the project: email and Google+ Hangout. It is worth mentioning that in
addition to this IRC was used throughout the project and
commercial tools like Flowdock were experimented with but not found to provide
enough benefit to be taken into active use.

\subsection{Limitations}

% This could be put elsewhere but it's not a limitation of the research itself:
%Some students might put their personal life before school projects, which can be
%seen in participation and communication activity. Furthermore, students might
%find it difficult to ask others to perform better or participate more actively
%as they do not have similar authority as managers have in industry.

%Alternatively in industry, employees are usually quite well monitored by the
%management, which might have a positive impact the employees' way to
%communicate.

We researched a student project so the setup is different from the industry. In
collecting the quantitative data we relied on the students to provide the data
instead of for example having access to the email server that emails were sent
through. But this should not skew the results because students reported that
they willingly gave the information as we only requested access to the headers.
We only researched a single case from single year, and we had only three teams,
so we had quite limited amount of data. The validity of our data would increase
by having observations from multiple projects and years.

As mentioned when describing the course in Course Description one \Canadian{}
student per team had a special function that did not fit within the parameters
of Scrum. As they were selected from the pool of the students furthest in their studies and
with the most industrial experience, we find it likely the special role that
u10 held did not influence the fact that the developer emerged as the leader in
light of communication. However, we can not be certain about this. For this
reason when repeating the experiment we should do it in a way that this problem
does not exist. In this case the Scrum Master was shared between the three teams.
If a scrum master is used per team, then that person can be used as the contact
person or then the instructor should communicate with the team instead of an
individual like Scrum expects.

\section{Conclusions and Future Work} \label{sec:conclusion}

By looking at both the main textual asynchronous communication practice, email,
and the main verbal synchronous communication practice, standups, we both
verified earlier observations on GSD projects and provided new findings on top
of them. As mentioned in previous work, communication is a key challenge and
this case study did not dispute that. However, by reflecting and acting on the
information, the students were able to change their communication practices in a
way that increased their satisfaction for the practices. Overall, students
were more satisfied for used communication practices as they decreased the
amount of sent emails.

In this as well as previous projects students work patterns did not follow the
regular office hours or days. The practice of having standups had the side
effect of clustering the work on code to the same days as well and making sure
that students worked on the project during both weeks of the 5 sprints. Still it
was visible that students seemed to work more as the standups came closer,
and they had to report what they had done. As such, standups are a good agile
practice adapted to GSD that pace the work of students.

In contrast to what Scrum expects, by looking at the communication for each team
during sprints, we identified an emergent leader with one of the three teams.
Based on the observation we thought to suggest that showing this information to
teams after a sprint would be useful for reflection.

We expect to organize the course again in fall 2013, which provides us the
opportunity to collect more data on this course setup and base our results on
multiple years. One option could be inspecting how much students spend time for
discussing using videoconferencing versus using asynchronous methods such as
email, and measuring how does that affect the overall satisfaction or the
results of the sprints.

Finally we want to thank all the students who participated on the course.
Without their consent to arrange standups late in the evening or early in the
morning, organizing a distributed course using Scrum between sites located ten
time zones apart would not have been possible. Hopefully they are now better
prepared for the realities in the industry.

% -------------- Bibliography ----------------------
\iftoggle{coins}{}{
\clearpage
}

\bibliographystyle{unsrtnat}

% Change title of bibliography
%\renewcommand{\refname}{References}
%\renewcommand{\bibname}{References}  % For book, report-style

\bibliography{sources}

\end{document}